\documentclass[aps,pre,twocolumn,superscriptaddress,showpacs]{revtex4-1}
\usepackage{graphicx}
\usepackage[mathlines]{lineno}
\usepackage{amsmath}
\usepackage{amssymb}
\usepackage{times}
\usepackage[dvips]{color}
\usepackage{mathtools}
\usepackage{bm}

\begin{document}

\title{Small-angle scattering from multi-phase fractals}
\author{A. Yu. Cherny}
\affiliation{Joint Institute for Nuclear Research, Dubna
141980, Moscow region, Russian Federation}

\author{E. M. Anitas}
\affiliation{Joint Institute for Nuclear Research, Dubna
141980, Moscow region, Russian Federation} \affiliation{Horia Hulubei National Institute
of Physics and Nuclear Engineering, RO-077125 Bucharest-Magurele, Romania}

\author{V. A. Osipov}
\affiliation{Joint Institute for Nuclear Research, Dubna
141980, Moscow region, Russian Federation}

\author{A. I. Kuklin}
\affiliation{Joint Institute for Nuclear Research, Dubna
141980, Moscow region, Russian Federation}\affiliation{Laboratory for advanced studies of membrane proteins, Moscow Institute of Physics and Technology, 141700 Dolgoprudniy, Russian Federation}

\date{\today}
\begin{abstract}
Small-angle scattering (SAS) intensities observed experimentally are often characterized by the presence of successive power-law regimes with various scattering exponents whose values vary from -4 to -1. This usually indicates multiple fractal structures of the sample characterized by different size scales. The existing models explaining the crossover positions (that is, the points where the power-law scattering exponent changes) involve only one contrast parameter, which depends solely on the ratio of the fractal sizes. Here, a model that describes SAS from a multi-phase system with a few contrast parameters is described, and it is shown that the crossover position depends on the scattering length density of each phase. The Stuhrmann contrast variation method is generalized and applied to experimental curves in the vicinity of the crossover point beyond the Guinier region. The contrast variation is applied not to the intensity itself but to the model parameters, which can be found by fitting the experimental data with the suggested interpolation formula. The model supplements the existing two-phase models and gives the simple condition of their inapplicability: if the crossover point depends on the contrast then a two-phase model is not relevant. The developed analysis allows one to answer the qualitative question of whether one fractal `absorbs' another one or they are both immersed in a surrounding homogeneous medium like a solvent or solid matrix. The models can be applied to experimental SAS data where the absolute value of the scattering exponent of the first power-law regime is higher than that of the subsequent second power-law regime, that is, the scattering curve is `convex' near the crossover point. As is shown, the crossover position can be very sensitive to contrast variation, which influences significantly the length of the fractal range.
\end{abstract}

\maketitle                        

\section{Introduction}

One of the most effective and reliable experimental method for investigating fractal structures \cite{mandelbrot82,gouyet96} of materials at nanoscales is small-angle scattering \cite{gf55,gk82,fs87} that measures the differential cross section of an irradiated sample \emph{versus} the scattering vector magnitude $q$ [$q = (4\pi/\lambda)\sin\theta$, where $\theta$ is half the scattering angle, and $\lambda$ is the wavelength of the incident radiation]. SAS allows one to obtain the edges of fractal region in reciprocal space and the fractal dimension \cite{bale84,schmidt86,martin87,teixeira88,schmidt91}. The borders of linear dependance of the intensity in a double logarithmic scale shows the edges of the fractal region, and its slope $\tau$, being the power-law exponent of SAS intensity, yields the fractal dimension: $I(q)~\propto~q^{-\tau}$, where
\begin{equation}\label{exptau}
\tau = \begin{cases}
   D, &\text{for mass fractals},\\
   6-D_{\mathrm{s}}, &\text{for surface fractals}.
  \end{cases}
\end{equation}
Here the notations $D$ and $D_{\mathrm{s}}$ are adopted for the fractal dimension of mass and surface fractals, respectively. The value of the fractal dimension can lie within the ranges $0<D<3$ and $2<D_{\mathrm{s}}<3$ for mass and surface fractals, respectively. These restrictions lead to a simple physical interpretation of experimental data: if $\tau<3$ then the measured sample is a mass fractal, if $3<\tau<4$ then it is a surface fractal. Note that this rule of thumb should be used with caution, because the appearance of the power-law behaviour can arise not only due to the fractal structure of the irradiated sample. For instance, the scattering intensity from spatially and rotationally uncorrelated identical disks yields the exponent equal to $-2$. Another example is an `occasional' power-law behaviour in a quite narrow momentum range owing to polydispersity. This is a general and well-known problem of the small-angle scattering method, which cannot give us a well-defined and unambiguous structure of samples in real space. In order to resolve this general ambiguity and reveal the sample structure in real space, one needs to involve additional information from other experimental techniques or by knowing the conditions under which the system has been created.

SAS data (X-ray or neutron) often show a succession of power-law regimes whose scattering exponents are taking arbitrarily values from $-4$ to $-1$ \cite{ehrburgerdolle01,schaefer07,kravchuk09,schneider10,singh11,schneider12,golosova12}. On a double logarithmic scale, the scattering intensity looks like connected straight-line segments. This type of behavior is usually associated with two-phase multilevel structures and the structural characteristics are revealed most often by using Beaucage model \cite{beaucage95}. However, experimental SAS papers have reported the dependence of the scattering intensities on the scattering length densities of surrounding solvent or solid matrix, in which the studied structures are immersed. In particular, the crossover positions $q_{\mathrm{c}}$ (that is, the points where the power-law scattering exponent changes) can depend on the contrast value  \cite{lebedev05}. This indicates a multi-phase structure of the sample, and thus a method capable of extracting the structural properties about each phase is needed.

Let us explain the problem in more detail. The multi-level structures suggested by Beaucage \cite{beaucage95} assumes that particles composing a multi-level system are complex structures themselves within their size-scale. An example is a mass fractal with fractal dimension $D_1$ composed of the structural units, where each of them is another mass fractal with the dimension $D_2$. Then the scattering intensity on a double logarithmic scale shows the `knee' behaviour, where one straight line with the slope $-D_1$ passes into another straight line with the slope $-D_2$ at the crossover point $q_{\mathrm{c}}$. This model is \emph{two-phase}, that is, it assumes that primary units of the above construction are homogeneous with a constant scattering length density and all the construction is embedded in a solid matrix with another constant scattering length density. Hence, there is only one contrast parameter (that is, the difference between the scattering length densities), and the scattering intensity is proportional to the squared value of the contrast, but its form remains the same with the contrast variation. In particular, the value of the crossover position  is independent of the contrast value but determined only by the ratio of the fractal sizes.

On the other hand, multi-phase materials are ubiquitous in nature, and one can expect that the presence of more than two phases could be even more relevant. In a recent paper \cite{chernyJPCS12}, the crossover between Porod and mass fractal regimes has been explained for multi-phase systems. Here, we extend these results and develop a model which describe SAS from \emph{multi-phase} system with \emph{a few contrast parameters}. As is known, the method of contrast variation \cite{stuhrmann70} is successfully used in the X-ray and neutron scattering for investigating multi-phase systems. This powerful method studies the intensity at zero momentum transfer and radii of gyration, and thus it is applicable only in the Guinier region. However, the Guinier region often lies beyond the resolution of experimental device. We suggest an extension of this method and show how it can be applied to experimental curves in the vicinity of the crossover point \textit{beyond the Guinier region}. The contrast variation is applied to the model parameters, which can be found by fitting the experimental data with the simple interpolation formula.

There are essential differences between the model suggested here and the Beaucage model. First, the structure of material in real space is assumed to be different. \label{options} In contrast to the real-space structure of the Beaucage model described above, we consider two possible different structures (see Fig.~\ref{fig:fig3} below):\\
Type \emph{I}. A fractal with the scattering length density $\rho_{2}$ and dimension $D_{2}$ is \emph{embedded in a bigger fractal} with the density $\rho_{1}$ and dimension $D_{1}$, and they are immersed in a homogeneous medium (solvent or solid matrix) with density $\rho_{0}$.\\
Type \emph{II}. Two \emph{non-overlapping fractals} with the dimensions $D_{1}$ and $D_{2}$ and scattering length densities $\rho_{1}$ and $\rho_{2}$ are immersed in a homogeneous medium with density $\rho_{0}$.\\
Second, the suggested model is applicable if the scattering intensities are `convex' in the vicinity of the crossover point on a double logarithmic scale, that is, a steeper line at small values of momentum transfer passes into a flatter line at the crossover point $q_{\mathrm{c}}$. This means that the fractal with higher power-law exponent dominates at small values of momentum, while its contribution to the scattering intensity is suppressed at large values of momentum. Here we give a recipe how in practice to distinguish between the structures of type \emph{I} and \emph{II} and determine the densities $\rho_{1}$ and $\rho_{2}$ by using the contrast variation method.

We confirm our conclusions by doing numerical calculations with randomly oriented and uniformly distributed three-phase systems containing balls and mass fractals, embedded in a solvent or solid matrix. The basic model for the numerical calculations of mass fractals is the generalized Cantor fractals with controllable fractal dimension \cite{chernyJACR10,chernyPRE11}.

The paper is organized as follows. In Sec.~\ref{sec:theory}, we introduce some notations and emphasize some important issues about SAS from two- and three-phase systems. In Sec.~\ref{sec:results} we derive a simple fitting formula describing SAS from multi-phase systems and show how the contrast variation method can be used in the vicinity of the crossover point. The crossover position is estimated through the model parameters, and the obtained results are discussed. In Sec.~\ref{sec:application} we show numerically that the developed model describes well the transition from Porod to mass fractal regimes.

\section{Small-angle scattering from two- and three-phase systems}
\label{sec:theory}

Let us consider a sample, consisting of microscopic scatterers with the scattering length $b_j$. Then the differential cross section of elastic scattering is given by \cite{fs87} $\mathrm{d}\sigma/\mathrm{d}\Omega=|A_{\mathrm{t}}(\bm{q})|^2$, where
\begin{equation}\label{scatampl}
A_{\mathrm{t}}(\bm{q})\equiv \int_{V'} \rho_\mathrm{s}(\bm{r}) e^{i \bm{q}\cdot\bm{r}}\mathrm{d}^3 r
\end{equation}
is the total scattering amplitude and $V'$ is the total volume irradiated by the incident beam. Here, the scattering length density $\rho_\mathrm{s}(\bm{r})$ is defined with the help of Dirac's $\delta$-function: $\rho_\mathrm{s}(\bm{r})=\sum_j b_j\delta(\bm{r}-\bm{r}_j)$, where $\bm{r}_j$ are the spatial positions of the scatters.

Note that a constant shift of the scattering length density in the overall sample is important only for very small values of wave vector $q$, unattainable with the SAS technique. This property is convenient to exclude the `background' density of the sample.

\subsection{Two-phase systems}

We consider first a two-phase sample composed of rigid homogeneous objects with the scattering length density $\rho$ immersed in a solid matrix of the density $\rho_0$. Therefore, by subtracting the `background' density $\rho_0$, we can consider the system as if the objects were `frozen' in a vacuum and had the density $\Delta\rho =\rho - \rho_{0}$. It is called scattering contrast. Besides, the objects are supposed to be of the same volume and shape, and their spatial positions and orientations are uncorrelated. Then the scattering intensity (that is, the cross section per unit volume of the sample) is given by
\begin{equation}
I(q) \equiv\frac{1}{V'}\frac{\mathrm{d}\sigma}{\mathrm{d}\Omega} = n|\Delta\rho|^{2} V^{2}\left\langle \left|F(\bm{q})\right|^{2}\right\rangle,
\label{eq:intensitygeneral}
\end{equation}
where $n$ is the object concentration in the sample, $V$ is the volume of each object, and $F(\bm{q})$ is the normalized scattering amplitude of the object
\begin{equation}
F(\bm{q})=\frac{1}{V}\int_{V}e^{-i\bm{q}\cdot\bm{r}}\mathrm{d}\bm{r},
\label{eq:formfactorgeneral}
\end{equation}
obeying the condition $F(0)=1$. It is also called the normalized form factor. The brackets $\left\langle \cdots \right\rangle$ stand for the mean value given by the ensemble averaging over all orientations of the objects. If the probability of any orientation is the same, then the mean value can be calculated by averaging over all directions $\bm{n}$ of the momentum transfer $\bm{q}=q \bm{n}$, that is, by integrating over the solid angle in the spherical coordinates ${q}_{x}=q \cos\varphi \sin\vartheta$, ${q}_{y}=q \sin\varphi \sin\vartheta$ and ${q}_{z}=q \cos\vartheta$
\begin{equation}
\langle f(q_x,q_y,q_z)
\rangle\equiv\frac{1}{4\pi}\int_{0}^{\pi}\mathrm{d}\vartheta\sin\vartheta\int_{0}^{2\pi}\mathrm{d}
\varphi\,f(q,\vartheta,\varphi). \label{aver}
\end{equation}

\begin{figure}
\centering\includegraphics[width=.8\columnwidth,clip=true]{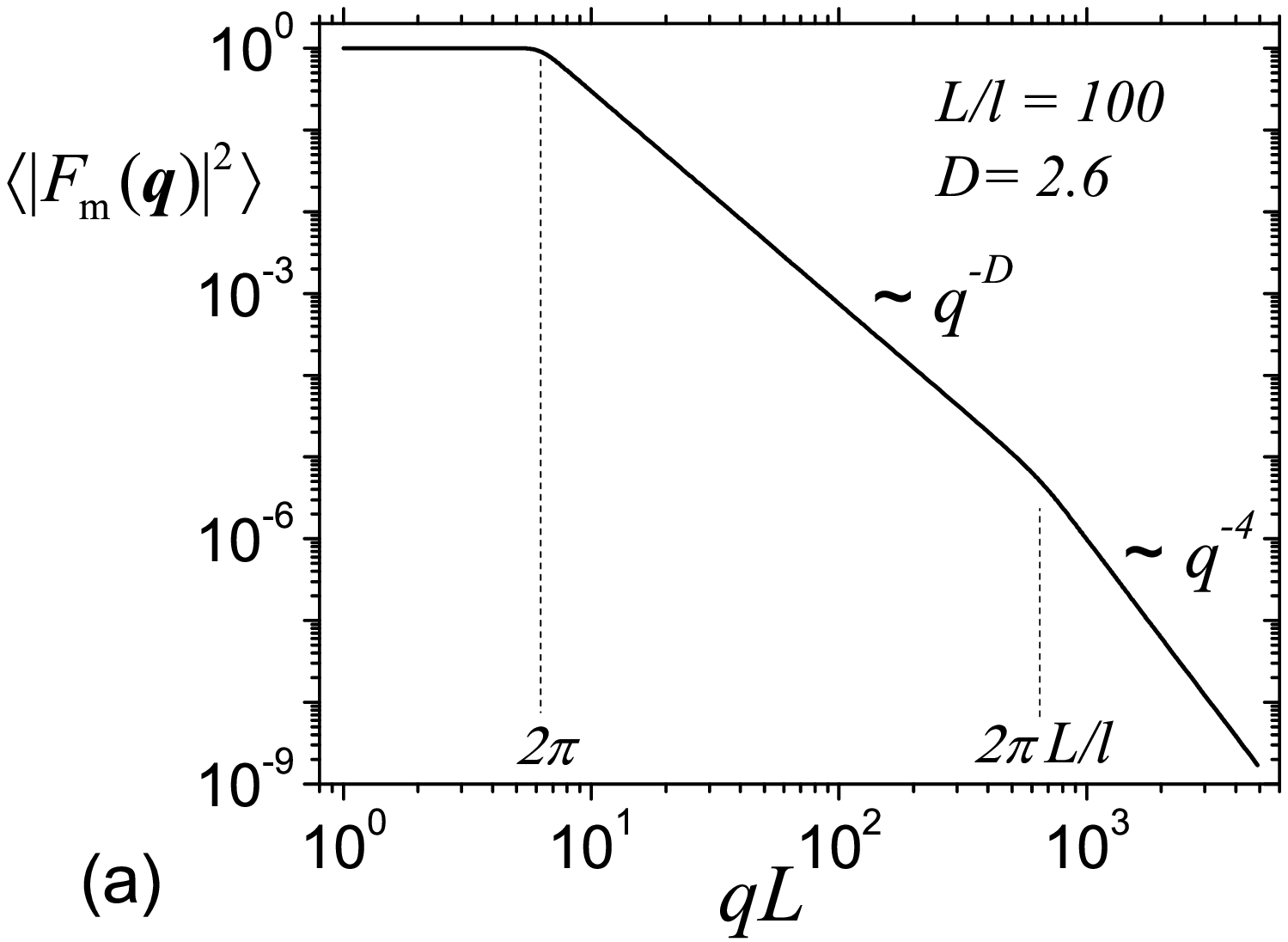}\\[.5 em]
\centering\includegraphics[width=.8\columnwidth,clip=true]{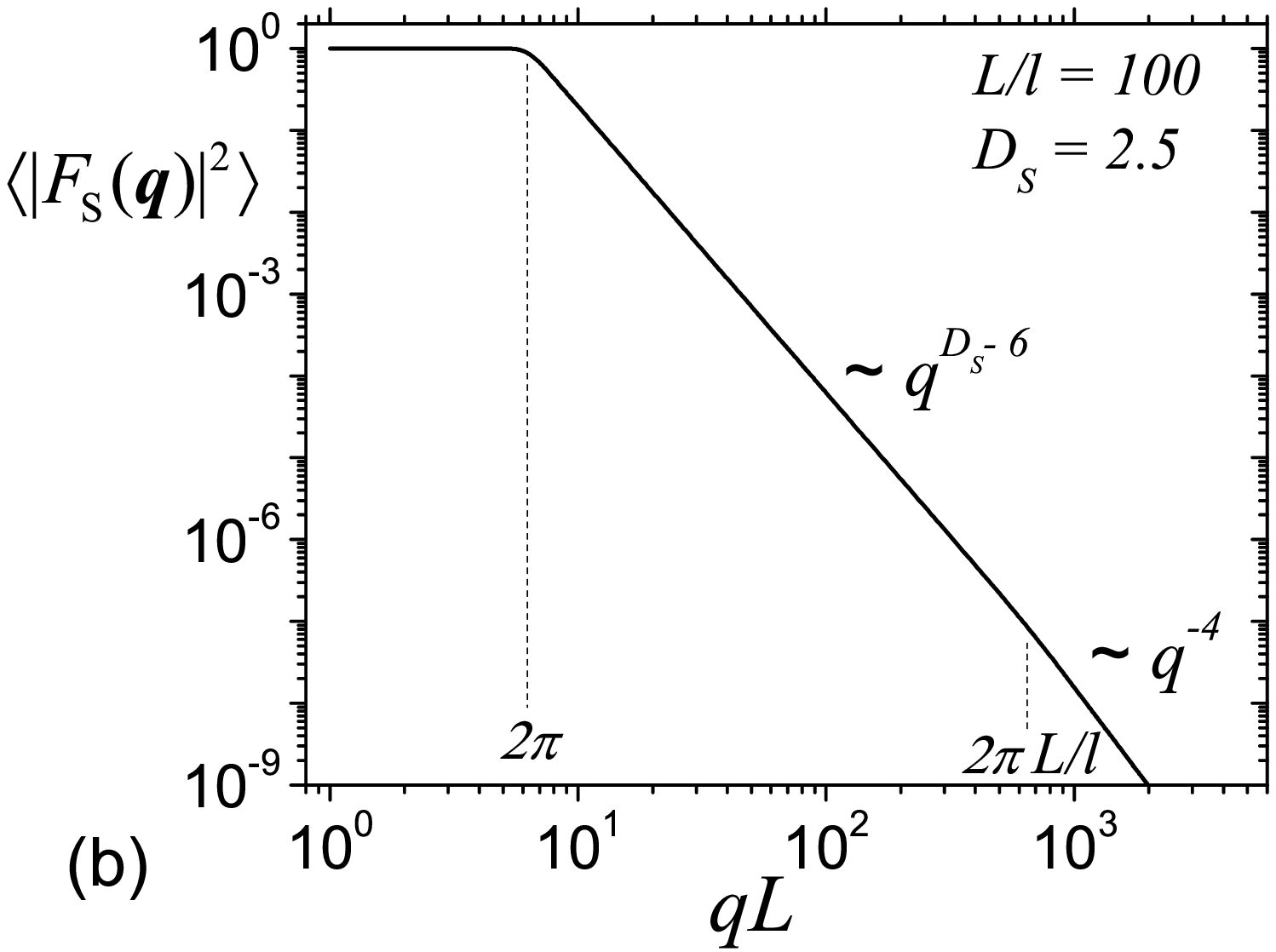}
\caption{\label{fig:fig1}A schematic representation of SAS from mass (a) and surface (b) fractals. The intensities show the three main ranges: Guinier (at small $q$), fractal (at intermediate $q$) and Porod (at high $q$). The characteristic lengths $L$ and $l$ are explained in the text.}
\end{figure}

If the object is a mass fractal of the total length $L$, composed of $p$ small structural units of the size $l$, then its normalized form factor can be estimated qualitatively by the formula
\begin{equation}
\langle|F_{\mathrm{m}}(\bm{q})|^{2}\rangle \simeq
\begin{dcases}
   1, &q \lesssim {2\pi}/{L},\\
   (q L/2\pi)^{-D}, &2\pi/L \lesssim q \lesssim {2\pi}/{l},\\
   \frac{1}{p}(ql/2\pi)^{-4}, &q \gtrsim 2\pi/l,
\end{dcases}
    \label{eq:fmass}
\end{equation}
see Fig.~\ref{fig:fig1}a. Here $p$ is of order $(L/l)^{D}$ in accordance with the definition of the fractal dimension, obeying the condition $0<D<3$.

A surface fractal can be constructed, say, from balls with different radii $r$, whose centers are spatially uncorrelated \emph{inside the fractal}. If the radii are taken at random with the distribution proportional to $1/r^{D_{\mathrm{s}}+1}$ at small $r$ and the parameter $D_{\mathrm{s}}$ obeys the condition $2<D_{\mathrm{s}}<3$ then one can show that $D_{\mathrm{s}}$ is indeed the fractal dimension of the total ball surfaces. Certainly, in practice, the restriction $l\leqslant r\leqslant L$ should be imposed, and the fractal properties appear only within this range. Now the qualitative estimating formula takes the form
\begin{equation}
\langle|F_{\mathrm{s}}(\bm{q})|^{2}\rangle \simeq
\begin{dcases}
   1, &q \lesssim {2\pi}/{L},\\
   (q L/2\pi)^{D_{\mathrm{s}}-6}, &2\pi/L \lesssim q \lesssim {2\pi}/{l},\\
   (L/l)^{D_{\mathrm{s}}-6}(ql/2\pi)^{-4}, &q \gtrsim 2\pi/l,
\end{dcases}
    \label{eq:fsurf}
\end{equation}
see Fig.~\ref{fig:fig1}b. We emphasize that here $L$ is of order of the largest radius of the balls but not the total fractal length, and $l$ is the smallest radius. Equations (\ref{eq:fmass}) and (\ref{eq:fsurf}) explicitly show that the scattering intensity in momentum space is characterized by three main ranges: Guinier at $q \lesssim {2\pi}/{L}$, fractal at $2\pi/L \lesssim q \lesssim {2\pi}/{l}$ and Porod at $q \gtrsim 2\pi/l$. Note that a simple three-dimensional set like a ball does not exhibit fractal behaviour at all; here the Porod range follows the Guinier range. Indeed, the fractal region `collapses' if we put formally $l=L$ in either of the relations (\ref{eq:fmass}) and (\ref{eq:fsurf}).

We make a few remarks about the scattering intensities (\ref{eq:fmass}) and (\ref{eq:fsurf}). First, the intensity in the Guinier range is actually parabolic: $I(q)\simeq I(0)(1-R_{\mathrm{g}}^2 q^2/3)$.
This parabolic behaviour of the intensity is ignored in the above estimations for the sake of simplicity. Second, the relation (\ref{eq:fmass}) is applicable at large momentum $q \gtrsim 2\pi/l$ if the minimal distance between the structural units composing the mass fractal is of order of their size. Otherwise, if the size is much smaller than the minimal distance then there exists a plateau (`shelf') between the end of the fractal range and the beginning of the Porod range \cite{chernyJACR10}. Third, the fractal region of the mass fractal (Fig.~\ref{fig:fig1}a) is determined by the maximal and minimal distances between the centers of the structural units, while the fractal region of the surface fractal (Fig.~\ref{fig:fig1}b) is determined by the biggest and smallest radii present in the fractal. For a mass fractal in the fractal region, the interparticle spatial correlations between the structural units play an important role, and the fractal behaviour can be understood by analogy with optics \cite{chernyPRE11}, while for a surface fractal, the spatial correlations between the structural units are not important for its fractal properties.

\subsection{Three-phase systems}

\begin{figure}
\centering\includegraphics[width=\columnwidth,clip=true]{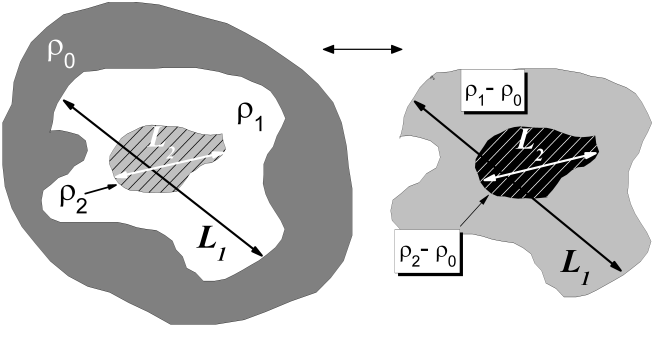}
\caption{\label{fig:fig2}A three-phase system composed of a homogeneous region 1 with the scattering length density $\rho_{1}$, which `absorbs' a region 2 with the density $\rho_{2}$. They together are embedded in a homogeneous medium (solvent or solid matrix) with the density $\rho_{0}$. The three-phase system is equivalent to the two-phase one in a vacuum with the contrast densities $\rho_{1}-\rho_{0}$ and $\rho_{2}-\rho_{0}$, respectively.
}
\end{figure}

The three-phase systems consist generically from homogeneous structures with the scattering length densities $\rho_{1}$ and $\rho_{2}$, immersed in a homogeneous medium with the density $\rho_{0}$. By subtracting the `background' density $\rho_{0}$, we arrive at the same structural units with the contrasts $\rho_{1}-\rho_{0}$ and $\rho_{2}-\rho_{0}$, respectively, in a vacuum (see Fig.~\ref{fig:fig2}). Effectively, this procedure allows us to reduce the three-phase system to the two-phase system.

Let us calculate the scattering intensity from a macroscopic number of the objects shown in Fig.~\ref{fig:fig2}, whose spatial positions and orientations are supposed to be uncorrelated. In accordance with the general relation (\ref{scatampl}), the intensity is given by $I(q)=n \langle|A(\bm{q})|^2\rangle$, where $A(\bm{q})$ and $n$ are the scattering amplitude of the objects and their concentration in the sample, respectively. The amplitude can be written down with the help of the normalized form factors $F_{1}(\bm{q})$ and $F_{2}(\bm{q})$ and volumes $V_{1}$ and $V_{2}$ of the space regions 1 and 2, respectively, shown on the left in Fig.~\ref{fig:fig2}. Indeed, it is given by a sum $A(\bm{q})=A_{\mathrm{1out}}(\bm{q})+A_{2}(\bm{q})$, where $A_{2}(\bm{q})=(\rho_{2}-\rho_{0})V_{2}F_{2}(\bm{q})$ is the scattering amplitude of the region 2, and $A_{\mathrm{1out}}(\bm{q})=(\rho_{1}-\rho_{0})[V_{1}F_{1}(\bm{q})-V_{2}F_{2}(\bm{q})]$ is the amplitude of the region that remains after excluding the region 2 from 1. Finally, we derive
\begin{equation}\label{int3phase}
I(q)=n\big\langle\big|(\rho_{1}-\rho_{0})V_{1}F_{1}(\bm{q})+(\rho_{2}-\rho_{1})V_{2}F_{2}(\bm{q})\big|^2\big\rangle.
\end{equation}

This expression can be written in the form \cite{chernyJPCS12}
\begin{equation}\label{int3phase_alp}
I(q)=I(0)\langle|\alpha F_{1}(\bm{q})+(1-\alpha)F_{2}(\bm{q})|^2\rangle,
\end{equation}
where we denote $I(0)=nV^2_{1}(\rho_{1}-\rho_{0})^2\alpha^{-2}$ and introduce by definition the `contrast parameter'
\begin{equation}
\alpha\equiv\left(1+\frac{\rho_{2}-\rho_{1}}{\rho_{1}-\rho_{0}}\frac{V_{2}}{V_{1}}\right)^{-1}.
\label{eq:alpha}
\end{equation}
This dimensionless parameter determines the relative contributions of the corresponding form factors, and thus it controls completely the form of the resulting scattering intensity, while the average scattering length density of the object $(\rho_{1}-\rho_{0})/\alpha$ determines the intensity at zero momentum.

\section{The interpolating formula and the generalized method of contrast variation}
\label{sec:results}

Let us analyze and further simplify the general relation (\ref{int3phase})  [or (\ref{int3phase_alp})] for the scattering intensity from the three-phase system. Suppose that the systems 1 and 2 shown in Fig.~\ref{fig:fig2} are mass or surface fractals, obeying the power-law behaviour in the corresponding fractal ranges in accordance with Eq.~(\ref{eq:fmass}) or (\ref{eq:fsurf}). We assume that the following conditions are satisfied:\\
\emph{(a)} their fractal ranges in momentum space overlap;\\
\emph{(b)} the corresponding \emph{power-law exponents are quite different} (recall that the exponents are directly connected to the fractal dimensions by the relation (\ref{exptau});\\
\emph{(c)} the dimensionless parameter $\alpha$ given by Eq.~(\ref{eq:alpha}) obeys the condition $|1-\alpha|\ll 1$ when $\tau_1>\tau_2$ or $|\alpha|\ll 1$ when $\tau_2>\tau_1$. Let us choose the first case without loss of generality.

Then the scattering intensity exhibits the `knee' convex structure on a double logarithmic scale. Indeed, at low momentum the first term in the r.h.s. of Eq.~(\ref{int3phase_alp}) dominates due to the condition \emph{(c)}. On the other hand, the first term is proportional to $F_1$, which decrease faster than $F_2$ according to the condition \emph{(b)}. At the crossover point $q_{\mathrm{c}}$, the both terms become approximately equal, and at high values of momentum the second terms dominates.

The physical picture can be simplified further if we notice that \emph{the interference terms}, proportional to $\langle F_{1}(\bm{q})F^*_{2}(\bm{q})\rangle + \langle F^*_{1}(\bm{q})F_{2}(\bm{q})\rangle$, \emph{can be neglected, provided one of the amplitude dominates}. This obvious statement can easily be proved even rigorously by means of the Cauchy-Schwarz inequality $|\langle F_{1}(\bm{q})F^*_{2}(\bm{q})\rangle|^2 \leqslant \langle|F_{1}(\bm{q})|^2\rangle \langle|F_{1}(\bm{q})|^2\rangle$, valid for arbitrary $\bm{q}$. Indeed, if for example, the second amplitude still dominates $\langle|F_{1}|^2\rangle \ll \langle|F_{2}|^2\rangle$ then $|\langle F_1 F^*_2\rangle| \leqslant \sqrt{\langle|F_{1}|^2\rangle}  \sqrt{\langle|F_{2}|^2\rangle }\ll \langle|F_{2}|^2\rangle$. Neglecting the interference terms in Eq.~(\ref{int3phase}) yields
\begin{equation}\label{int3phase_diag}
I(q)=(\rho_{1}-\rho_{0})^2 nV^2_{1}\langle|F_{1}(\bm{q})|^2\rangle+(\rho_{2}-\rho_{1})^2 nV^2_{2} \langle|F_{2}(\bm{q})|^2.
\end{equation}
Although this approximation is valid beyond the crossover position, the `crossover range' where the amplitudes are of the same order of magnitude always looks rather small on a double logarithmic scale. Therefore, we can use Eq.~(\ref{int3phase_diag}) even quite close to the crossover point $q_{\mathrm{c}}$. We show numerically in the next section that omitting the interference term is indeed a very good approximation.

\begin{figure}
\centering \includegraphics[width=.8\columnwidth,clip=true]{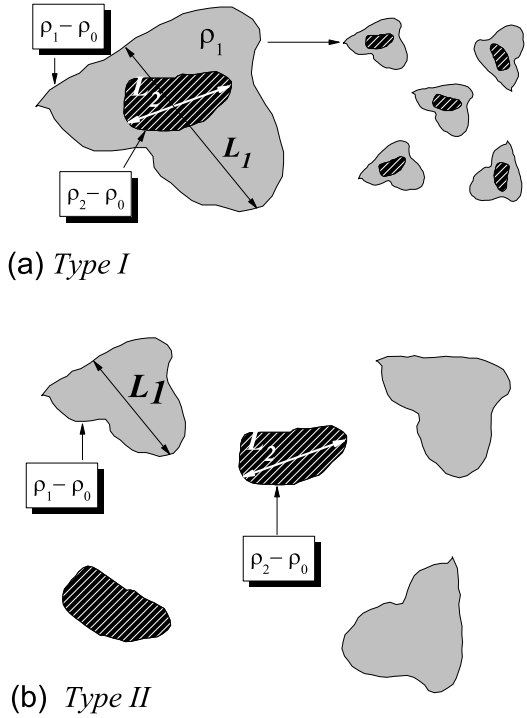}
\caption{\label{fig:fig3}Generic three-phase system of type \emph{I} and \emph{II}, described in detail in Introduction (see page~\pageref{options}). (a) The type \emph{I} structure is shown on the left panel: the region 2 is embedded in the region 1. The sample consists of a macroscopic number of these spatially uncorrelated structures, which are embedded in a solvent or solid matrix (right). The number of objects 1 and 2 is equal and hence their concentrations over the sample coincide. (b)  The type \emph{II} structures: the regions 1 and 2 do not overlap and their positions are uncorrelated in the sample. The concentrations of objects 1 and 2 are different in general.}
\end{figure}

Finally, the condition \emph{(a)} allows us to use the power-law behaviour of form factors in the fractal regimes given by Eqs.~(\ref{eq:fmass}) and (\ref{eq:fsurf}), and we derive from Eq.~(\ref{int3phase_diag})
\begin{equation}
I(q)=\frac{(\rho_{1}-\rho_{0})^{2}n V_1^2}{(qL_1/2\pi)^{\tau_1}}+\frac{(\rho_{2}-\rho_{1})^{2}n V_2^2}{(qL_2/2\pi)^{\tau_2}},
\label{eq:fittingintensity}
\end{equation}
 where the exponents $\tau_1$ and $\tau_2$ are given by Eq.~(\ref{exptau}). This is the interpolation formula for the scattering intensity from a macroscopic number of objects of type `fractal inside fractal', the type \emph{I} described in Introduction (see the left panel in Fig.~\ref{fig:fig3}).

One can consider a macroscopic number of non-overlapping fractals with concentrations $n_1$ and $n_2$, which all are spatially uncorrelated (system of type \emph{II} shown on the right panel in Fig.~\ref{fig:fig3}). The scattering intensity for this type of system can be obtained in the same manner as Eq.~(\ref{eq:fittingintensity})
\begin{equation}\label{int3phase_diagII}
I(q)=(\rho_{1}-\rho_{0})^2 n_1V^2_{1}\langle|F_{1}(\bm{q})|^2\rangle+(\rho_{2}-\rho_{0})^2 n_2V^2_{2} \langle|F_{2}(\bm{q})|^2.
\end{equation}
Finally, we have in the particular case when the both regions are in the scattering fractal regimes
\begin{equation}
I(q)=\frac{(\rho_{1}-\rho_{0})^{2}n_1 V_1^2}{(qL_1/2\pi)^{\tau_1}}+\frac{(\rho_{2}-\rho_{0})^{2}n_2 V_2^2}{(qL_2/2\pi)^{\tau_2}}.
\label{eq:fittingintensityII}
\end{equation}
Equations (\ref{eq:fittingintensity}) and (\ref{eq:fittingintensityII}) are very similar to each other, except for the density dependence in the second term. This difference can be used to distinguish between the types \emph{I} and \emph{II} experimentally.

The first and the second terms in the expression for scattering intensity [Eq.~ (\ref{int3phase_diag}) or (\ref{int3phase_diagII})] are of the same order of magnitude in the vicinity of the crossover point, and, therefore, equating these terms  yields the crossover position. Note that the form factors (\ref{eq:fmass}) and (\ref{eq:fsurf}) contain all the qualitative regions in the intensity curve: the Guinier, fractal and Porod regions, and thus this curve together with Eq.~(\ref{int3phase_diag}) or (\ref{int3phase_diagII}) allows us to obtain the crossover position for all the types of transitions: Guinier-fractal,  Guinier-Porod, fractal-fractal and fractal-Porod. For the fractal-fractal transition one can use directly Eqs. (\ref{eq:fittingintensity}) and (\ref{eq:fittingintensityII}) and obtain
\begin{equation} \label{eq:crossovermf}
q_{\mathrm{c}} \simeq 2\pi\left( \delta \frac{L_{1}^{\tau_{1}}}{L_{2}^{\tau_{2}}}
\right)^{1/(\tau_{2}-\tau_{1})},
\end{equation}
where $\delta$ is the dimensionless `contrast parameter'
\begin{equation}\label{eq:deltacontrast}
\delta=\frac{V_2^2}{V_1^2}\times\begin{cases}
\dfrac{(\rho_{2}-\rho_{1})^{2}}{(\rho_{1}-\rho_{0})^{2}},& \text{for type \emph{I}}\\
\dfrac{(\rho_{2}-\rho_{0})^{2}}{(\rho_{1}-\rho_{0})^{2}}\dfrac{n_{\mathrm{2}}}{n_{\mathrm{1}}},& \text{for type \emph{II}}.
\end{cases}
\end{equation}
Certainly, for the type \emph{I} structure (see Fig.~\ref{fig:fig3}), this parameter is connected simply with the `contrast parameter' $\alpha$ given by Eq.~(\ref{eq:alpha}) by the relation $\delta = (1-\alpha)^2/\alpha^2$.

The parameter $\delta$ plays an important role, because it determines the relative `weights' of the squared normalized form factors in Eqs.~(\ref{int3phase_diag}) and (\ref{int3phase_diagII}) [and correspondingly Eqs.~(\ref{eq:fittingintensity}) and (\ref{eq:fittingintensityII}) in the fractal ranges]. As a consequence, it governs the form of the scattering intensity. Note that the suggested model works only if $\delta\ll 1$ and simultaneously $\tau_1>\tau_2$ or inversely $\delta\gg 1$ and  $\tau_1<\tau_2$. Otherwise only one form factor dominates for all values of the momentum and there is no transition on the scattering curve at all. Within the model, one can obtain only the `convex' behaviour near the transition point on a double logarithmic scale.

The both approximations for the scattering intensities (\ref{eq:fittingintensity}) and (\ref{eq:fittingintensityII}) can be written in the form
\begin{equation}
I(q)=\frac{a_{1}}{q^{\tau_{1}}}+\frac{a_{2}}{q^{\tau_{2}}},
\label{eq:fitgen}
\end{equation}
where we put by definition
\begin{align}
a_{1}&=(2\pi/L_{1})^{\tau_{1}}n_1 V_1^{2}(\rho_{1}-\rho_{0})^{2}, \label{eq:acoefs1}\\
a_{2}&= (2\pi/L_{2})^{\tau_{2}}V_2^{2}\times\begin{cases}
n_1(\rho_{2}-\rho_{1})^{2},& \text{for type \emph{I}},\\
n_2(\rho_{2}-\rho_{0})^{2},& \text{for type \emph{II}},
\end{cases}
\label{eq:acoefs2}
\end{align}
and the exponents $\tau_1$ and $\tau_2$ are related to the corresponding fractal dimensions by Eq.~(\ref{exptau}).

One can prepare a number of samples, which differ only in the density $\rho_{0}$ of solvent or solid matrix, and measure their SAS intensities; this is the method of contrast variation, initially suggested by Stuhrmann \cite{stuhrmann70}. Fitting formula (\ref{eq:fitgen}) in conjunction with equations (\ref{eq:acoefs1}) and (\ref{eq:acoefs2}) suggests the extension of this method when it is applied not for the scattering intensity itself but for the coefficients $a_1$ and $a_2$. Indeed, fitting each intensity curve with Eq.~(\ref{eq:fitgen}) yields the values of dimensions $D_1$ and $D_2$ (through the exponents $\tau_1$ and $\tau_2$) and the values of these coefficients at given values of $\rho_0$. Since $\sqrt{a_1}\sim|\rho_{1}-\rho_{0}|$, one can find the value of $\rho_1$ as the contrast match point in the plot $\sqrt{a_1}$ versus $\rho_0$ and determine the value of $n_1V_1^{2}/L_{1}^{\tau_{1}}$ as well. \emph{A structure of the type II exhibits the similar dependence of the coefficient $a_2$}, and the values of density $\rho_2$ and $n_1V_1^{2}/L_{1}^{\tau_{1}}$ can be found the same way. \emph{If $a_2$ is independent of $\rho_0$, we deal with a structure of the type I} and hardly can extract more information than the contrast value $|\rho_{2}-\rho_{1}|$ itself unless other methods are used.

Very often a priori information about the fractal system under investigation is known, mainly from other experimental techniques or by knowing the conditions under which the system has been created. One can then use the found coefficients like $n_1V_1^{2}/L_{1}^{\tau_{1}}$ to obtain the unknown feature. For example, if the concentration $n_1$ is known from a sample preparation, the size $L_1$ of the region 1 can be estimated from the border of the Guinier range (see Fig.~\ref{fig:fig1}). Then the fractal volume $V_1$ can be calculated as well.

Note that the coefficients $a_1$ and $a_2$ is given by the asymptotics of the functions $q^{\tau_{1}}I(q)$ and $q^{\tau_{2}}I(q)$, respectively, beyond the crossover regime. Therefore, their values can be found from a Kratky-like plot on a double logarithmic scale (see Fig.~\ref{fig:fig6} below).
\begin{figure}
\centering\includegraphics[width=.9\columnwidth,clip=true]{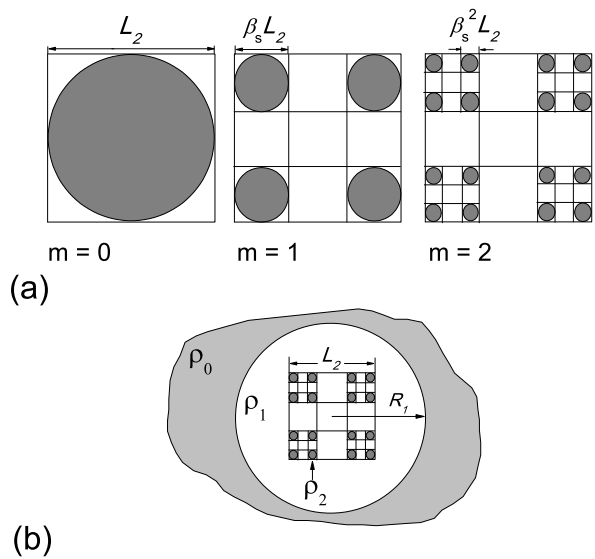}
\caption{\label{fig:fig4}(a) The $2D$ projection of zero, first and second iterations of the generalized Cantor fractal. Here we choose $\beta_{\mathrm{s}}=1/3$. (b) A specific structure of type \emph{I} (see Fig.~\ref{fig:fig2}): the Cantor mass fractal of the density $\rho_{2}$ (region $2$) is embedded in a ball of the density $\rho_{1}$ (region $1$).
}
\end{figure}

\section{An example: Porod to mass fractal transition}
\label{sec:application}

In order to confirm the above results, we consider an `exactly solvable' three-phase system of type \emph{I} (see Fig.~\ref{fig:fig2}): a mass fractal of the density $\rho_{2}$ (region $2$) embedded in a ball of the density $\rho_{1}$ (region $1$). A ball of radius $R_1$ can be considered as the limiting case of surface fractal with surface dimension $D_{\mathrm{s}1}=2$. This specific structure of type \emph{I} is shown in Fig.~\ref{fig:fig4}. In the range $q\gtrsim 2\pi/R_1$, the scattering intensity of ball is characterized by the Porod exponent, equal to $\tau_1=6-D_{\mathrm{s}1}=4$. Therefore, the scattering intensity of the system should exhibit the transition from the Porod range to that of mass fractal. The ball should be big enough to absorb the fractal completely. We choose $R_{1}=L_{2}\sqrt{3}/2$, where $L_2$ is the fractal size, and the centers of the ball and fractal coincide. The volume of the region $1$ is obvious: $V_1=4\pi R_1^3/3$, and its form factor is given by $F_1(q)=F_0(q R_1)$, where
\begin{equation}
F_{0}(z)=3(\sin z - z\cos z)/z^{3} \label{sphereff}
\end{equation}
is the form factor of ball of unit radius \cite{fs87}.

The mass fractal is the generalized Cantor fractal with controllable fractal dimension. Its construction and properties are described in detail in \cite{chernyJACR10}, see also \cite{chernyJSI10,chernyPRE11}. The initial ball of radius $L_2/2$ is replaced by $8$ balls of radius $\beta_{\mathrm{s}}L_2/2$, where $\beta_{\mathrm{s}}$ is the scaling factor, whose value lies between 0 and $1/2$. Then this `creating and scaling' operation is applied to each of the resulted balls again and again (see Fig.~\ref{fig:fig4}). The number $m$ of the operations is called fractal iteration. Then the total fractal volume is given by $V_2=(2\beta_{\mathrm{s}})^{3m}\pi L_2^3/6$. The fractal dimension of the volume, in the limit of infinite fractal iterations, is $D=-3\log 2/\log \beta_{\mathrm{s}}$. The dimension takes the values between $0$ and $3$ when the scaling factor varies from $0$ to $1/2$. This fractal belongs to the class of \textit{deterministic} fractals exhibiting the exact self-similarity. The form factor of the generalized Cantor fractal of $m$th iteration can be calculated analytically \cite{chernyJACR10}
\begin{equation}
F_{2}(\bm{q})=F_{0}(\beta_{\mathrm{s}}^{m}L_2 q/2)G_{1}(\bm{q})G_{2}(\bm{q})\cdots G_{m}(\bm{q}), \label{eq:gcfformfactor}
\end{equation}
where $F_{0}(z)$ is given by Eq.~(\ref{sphereff}) and we put by definition
\begin{equation}
G_m(\bm{q})=\cos(u_{m}q_{x})\cos(u_{m}q_{y})\cos(u_{m}q_{z})
\label{eq:generativefunction}
\end{equation}
with $u_{m}\equiv L_{2}(1-\beta_\mathrm{s})\beta_\mathrm{s}^{m-1}/2$.

Substituting the above form factors $F_1$ and $F_2$ in Eq.~(\ref{int3phase_alp}) yields the scattering intensity of the three-phase system. If the average is taken over the solid angles in accordance with Eq.~(\ref{aver}), we obtain not a smooth (like shown in Fig.~\ref{fig:fig1}) but oscillating curve where the oscillations are superimposed on the power-law regimes. The scattering curve like this is typical for \emph{deterministic} fractals \cite{chernyPRE11}. The smoothing effect, observed experimentally, arises as a consequence of randomness involved in the fractals. Indeed, in a physical system scatterers almost always have different sizes. Therefore, a more realistic description should involve size polydispersity. Here we consider an ensemble of three-phase fractals with different sizes taken at random. For the model, we choose the size distribution to be log-normal. It is characterized by the average size value and the size relative dispersion (see Refs. \cite{chernyJACR10,chernyPRE11} for details). Thus, the average in Eq.~(\ref{int3phase}) is taken both \emph{over angles and sizes}. Polydispersity obviously smears the intensity curves, and the oscillations become smoother \cite{chernyPRE11}.

\begin{figure}
\centering\includegraphics[width=0.87\columnwidth,clip=true]{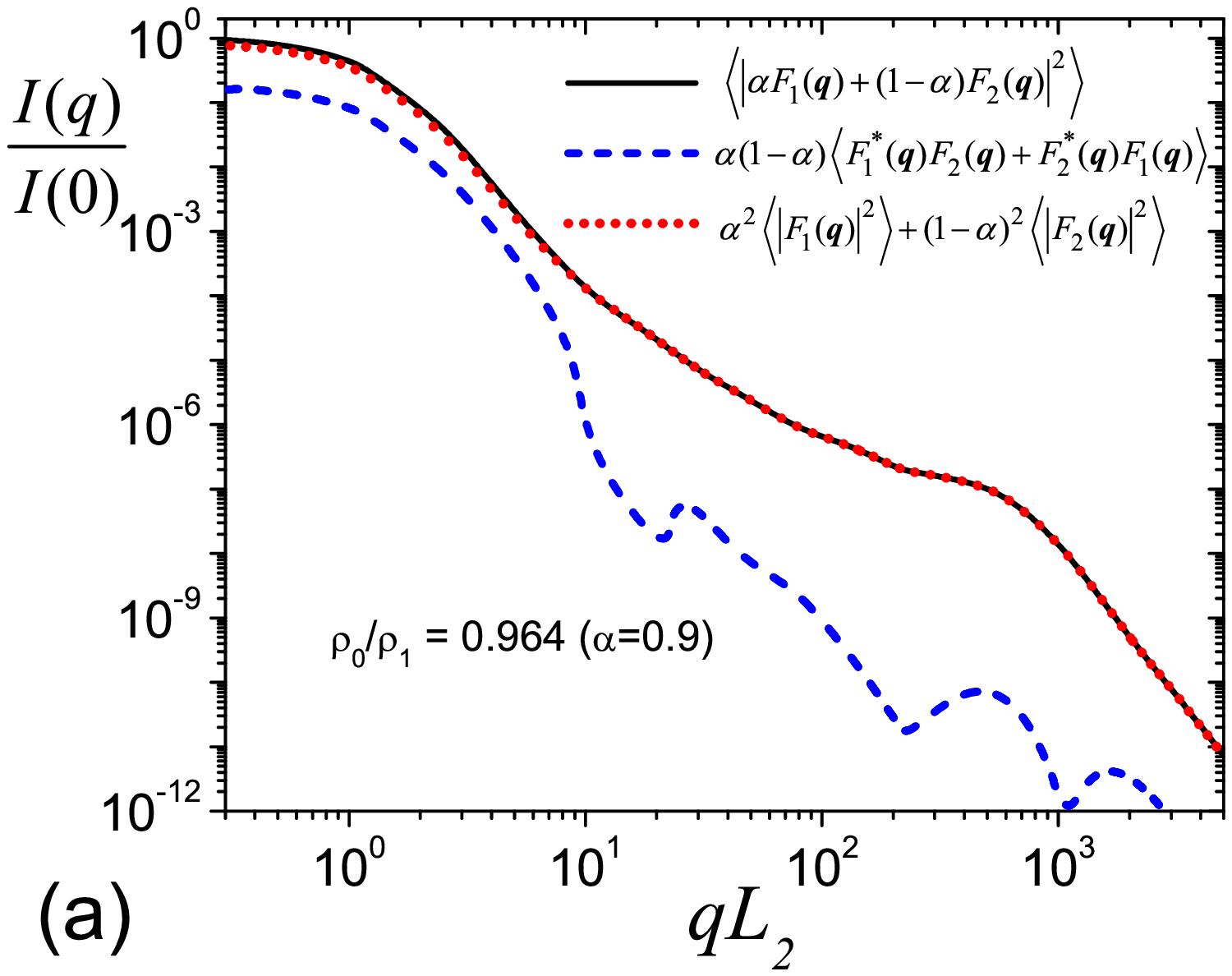}
\centering\includegraphics[width=0.87\columnwidth,clip=true]{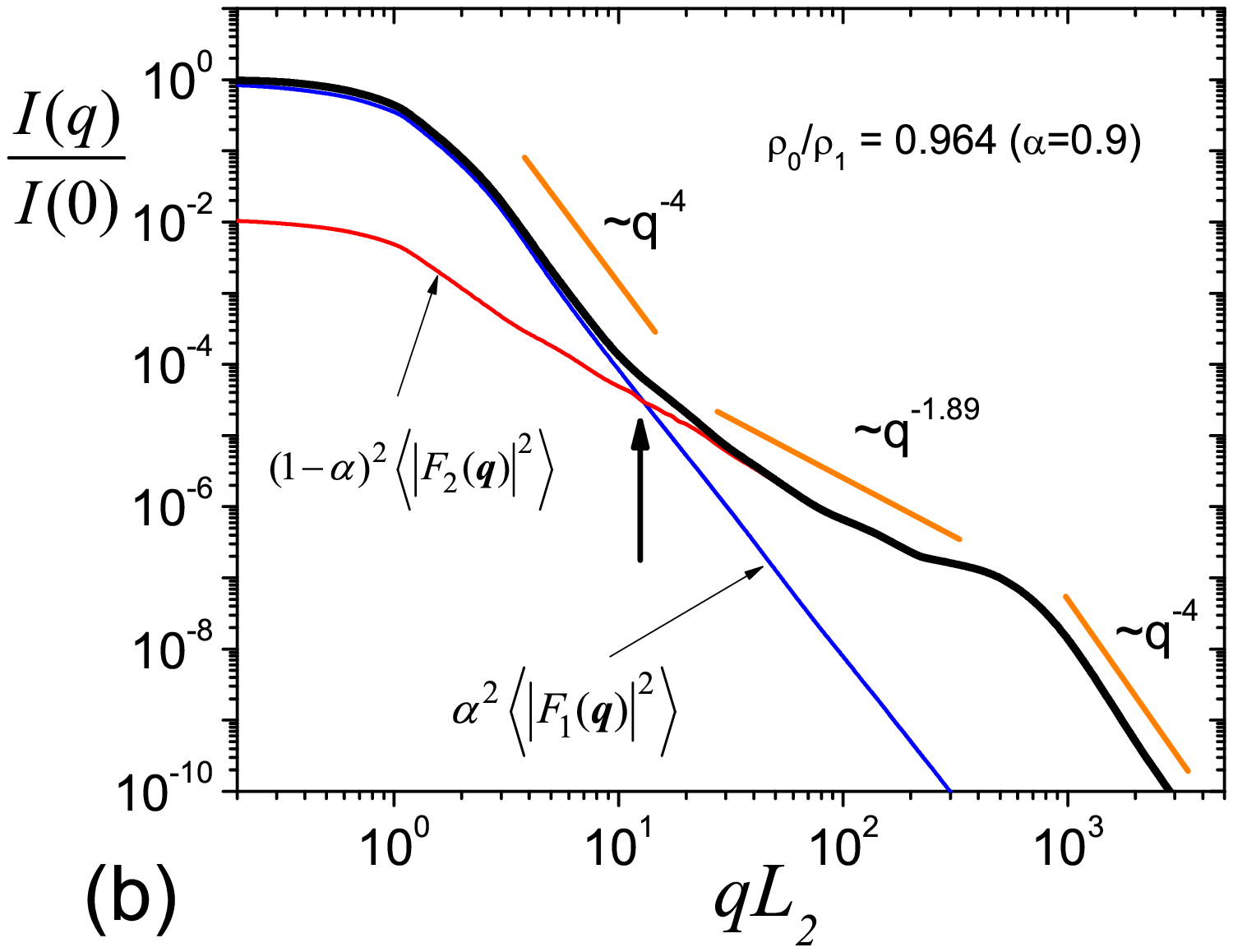}
\centering\includegraphics[width=0.89\columnwidth,clip=true]{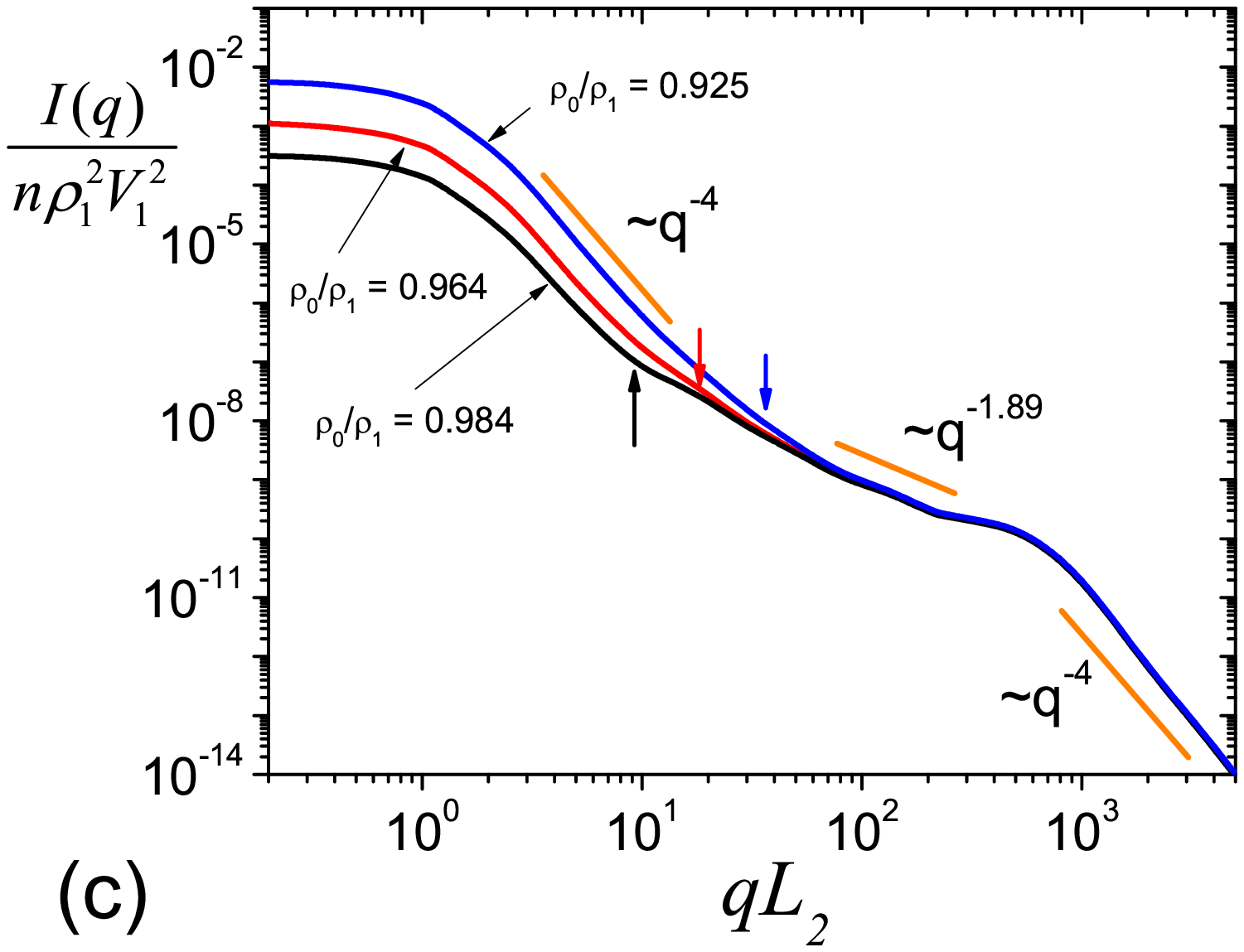}
\caption{\label{fig:fig5}The scattering intensities from polydisperse  three-phase mass fractal model of type \emph{I} (see Fig.~\ref{fig:fig3}), where the region 1 is a ball and the region 2 is the Cantor fractal. The parameters are described in the text. The vertical arrows represent the crossover position as estimated from Eq.~(\ref{eq:crossovermf}). (a) The contribution of various terms to the scattering intensity (\ref{int3phase_alp}): the full intensity (black solid line), the contribution of the `cross' terms (blue dashed line), and the contribution of the `diagonal' terms (red dotted line). The correlation terms are negligible on a double logarithmic scale. (b) The scattering intensity (black) and the contributions of the ball (blue) and mass fractal (red). Below the crossover point, the contribution of the mass fractal is negligible, while above it, the contribution of the ball is negligible. This is why the intensity is a convex function near the crossover point. (c) Contrast variation of the scattering intensity.}
\end{figure}
The results are shown in Fig.~\ref{fig:fig5}. For numerical simulations, we take the following values of the model parameters: the fractal iteration number $m=5$ and the scaling parameter $\beta_{\mathrm{s}}=1/3$ (which leads to the fractal exponent $\tau_{2}=1.89\ldots$, coinciding with the fractal dimension). It is convenient to measure the densities $\rho_2$ and $\rho_0$ in units of $\rho_1$ and the scattering intensity in units of $n_1\rho_1^2V_1^2$. We put $\sigma_{\mathrm{r}}=0.4$ for the relative dispersion of the log-normal distribution and choose $\rho_2/\rho_1=10$ for all the plots. Then the values of $\alpha$ depends on the ratio $\rho_0/\rho_1$. It can be seen clearly from Fig.~\ref{fig:fig5}a that the interference terms in Eq.~(\ref{int3phase_alp}) plays a minor role even in the Guinier range, where $\langle|F_{1}(\bm{q})|^2\simeq\langle|F_{2}(\bm{q})|^2\simeq 1$. This is due to the condition $|1-\alpha|\ll 1$.

\begin{figure}
\centering {\hspace*{.5cm}\includegraphics[width=.8\columnwidth]{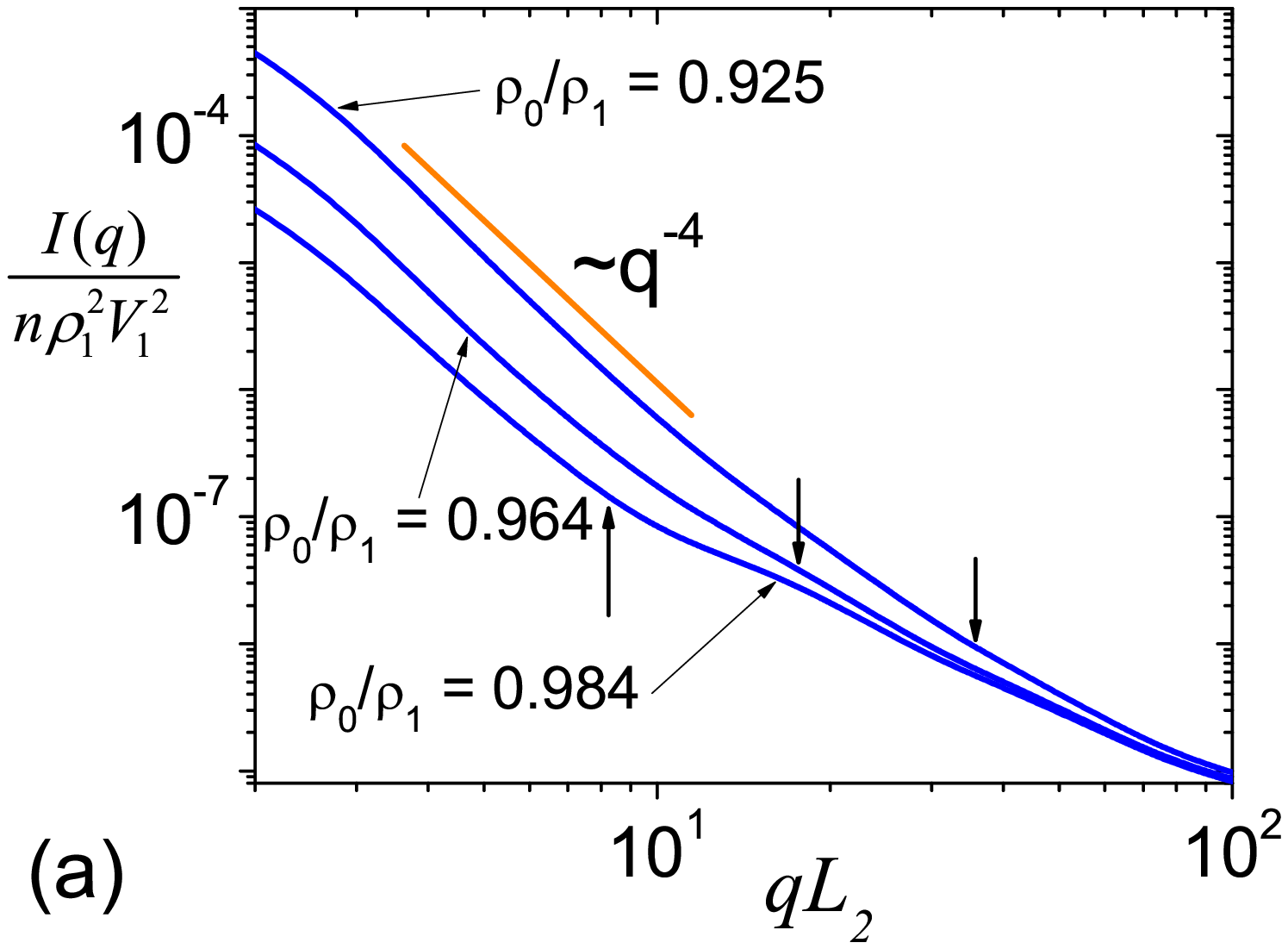}}
\centering \includegraphics[width=.85\columnwidth]{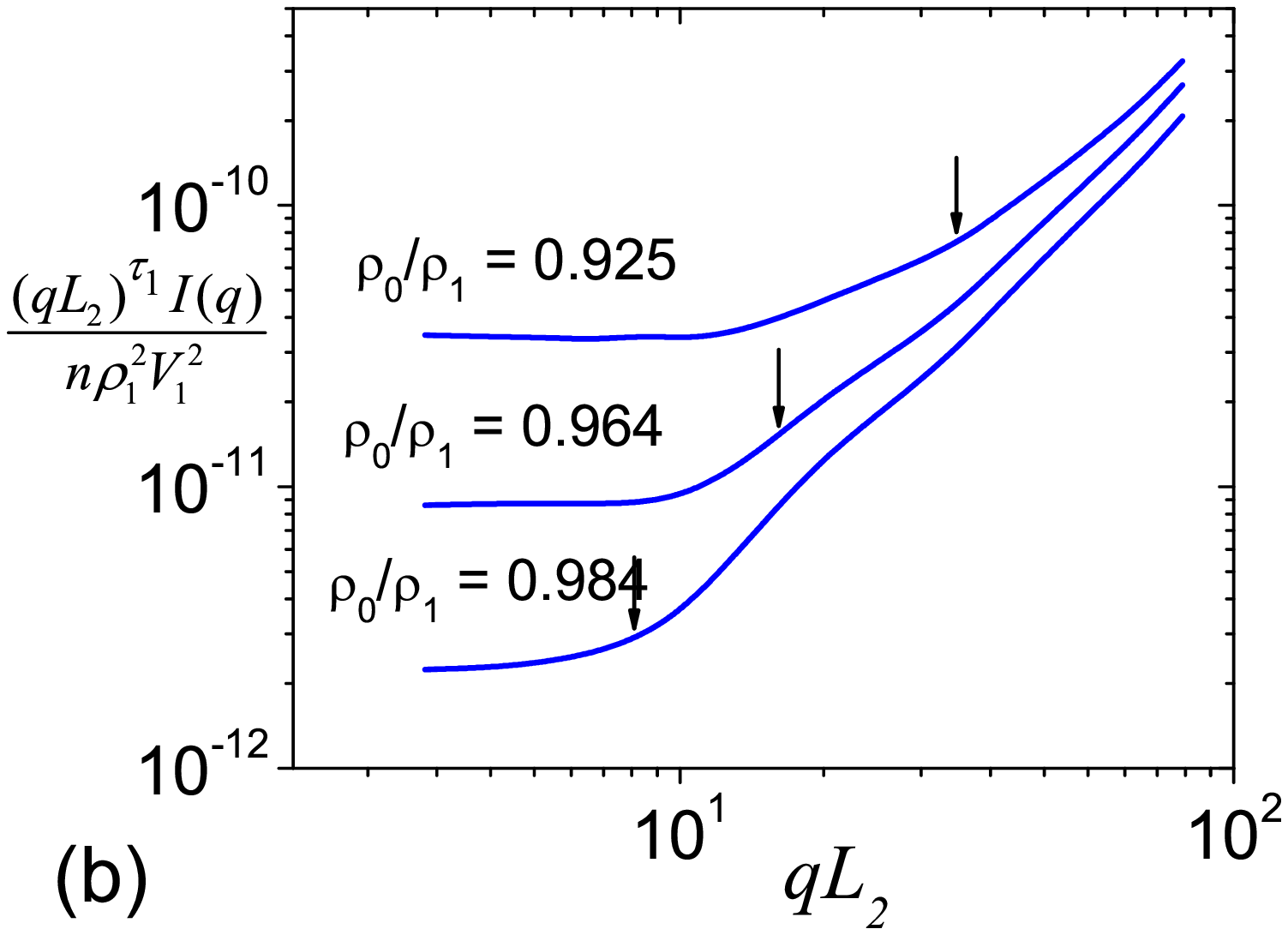}
\centering \includegraphics[width=.85\columnwidth]{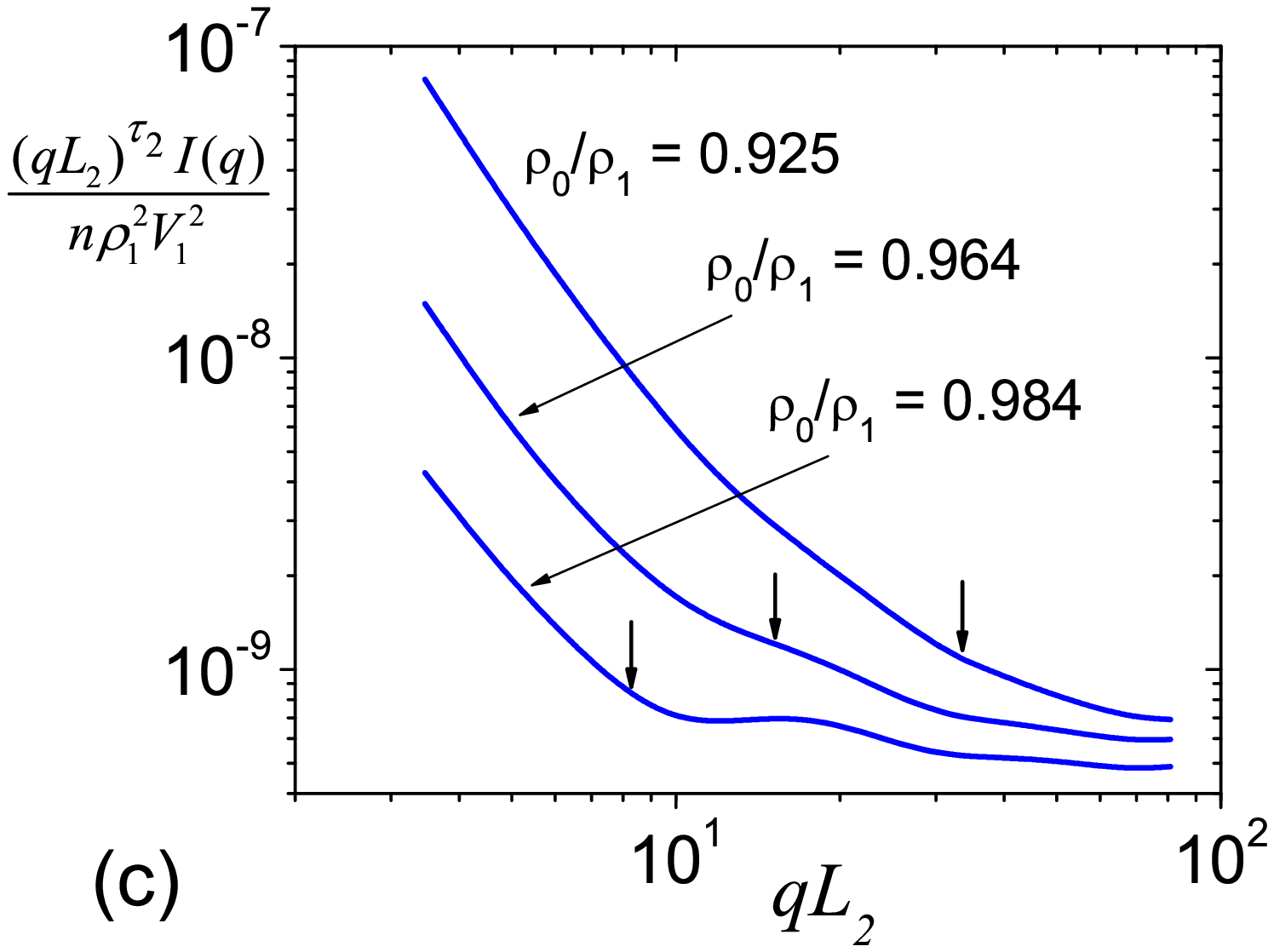}
\caption{\label{fig:fig6}The scattering intensity of Fig.~\ref{fig:fig5}c in a narrower $q$-range (a) and its Kratky-like plots for the exponent $\tau_{1}=4$ of the surrounding ball (b) and for the exponent $\tau_{2}=1.89\ldots$ of the generalized Cantor fractal (c). The vertical arrows show the positions of crossover as estimated from Eq.~(\ref{eq:crossovermf}). The contrast variation of the plateau values allows us to distinguish between the type \emph{I} and type \emph{II} systems, see Eqs.~(\ref{eq:fitgen})-(\ref{eq:acoefs2}).}
\end{figure}

Because of instrumental limitations, experimental SAS data may contain only a part of the whole scattering curve. The most common case is a curve showing only the transition between two power-law regimes, while the Guinier region and the Porod region at high $q$ are usually missing. Nevertheless, one can exploit the interpolation formula (\ref{eq:fitgen}) even in this case.

We remark some features of the model scattering intensities. First, the crossover position (\ref{eq:crossovermf}) can be very sensitive to contrast variation, which influences significantly the length of the fractal range (see Fig.~\ref{fig:fig5}c). While for $\rho_0/\rho_1=0.984$ the length of fractal range is big enough, for $\rho_0/\rho_1=0.925$ it becomes so small that once the intensity is obtained experimentally, this range can hardly be interpreted as a `true' fractal behaviour of the system. This suggests the necessity of the contrast variation experiments to reveal the structure of multi-phase system in real space. Second, at the end of the fractal range, one can see a `shelf' \cite{chernyJACR10}, which is nothing else but the Guinier range of the smallest structural units composing the fractal (the balls in our model). Such a behaviour is explained fairly well within the Beaucage model.

The transition of the scattering intensity from one regime to another one can be more clearly observed when we represent the contrast variation data in a Kratky-like plot. At low and high $q$-values in Figs.~\ref{fig:fig6}b and \ref{fig:fig6}c, respectively, one can clearly see plateaus corresponding to the coefficients $a_1$ and $a_2$ in Eq.~(\ref{eq:fitgen}), to which the Stuhrmann method of contrast variation  can be applied. The further we are, in the plateau region, from the crossover point, the better the approximation (see the previous section). The oscillations observed beyond the plateau regions damp with increasing the polydispersity.

\section{Conclusion}

We explain the origin and positioning of the crossover between successive power-law regime in SAS experimental data for multi-phase fractal systems of type \emph{I} and \emph{II}, see Fig.~\ref{fig:fig3}. The crossover positions and the contributions of the different structural levels to the total scattering intensity are controlled by the effective dimensionless `contrast parameter' $\delta$ given by Eq.~(\ref{eq:deltacontrast}), which depends on the relative values of the scattering length density of each phase and their volumes and concentrations.

A simple estimation of the crossover position (\ref{eq:crossovermf}) is obtained. As is shown numerically in Sec.~\ref{sec:application}, the crossover position can be very sensitive to contrast variation, which \emph{influences significantly the length of the fractal range}. The developed analysis is applicable for the scattering intensities that are `convex' in the vicinity of the crossover point on a double logarithmic scale and exhibits variation of the crossover point with contrast variation.

From a practical point of view, the main result of this paper is Eqs.~(\ref{eq:fitgen})-(\ref{eq:acoefs2}). They can be used to fit experimental SAS data and distinguish between the type \emph{I} system (one fractal absorbs the other fractal) and the type \emph{II} system (non-overlapping fractals) as described in Sec.~\ref{sec:results}. Once experimental intensity curves for a number of values of $\rho_0$ is available, one can fit the data with Eq.~(\ref{eq:fitgen}) and obtain the exponents $\tau_1$ and $\tau_2$ and the coefficients $a_1$ and $a_2$ as a function of the surrounding density $\rho_0$. Then plotting  $\sqrt{a_{1}}$ versus $\rho_{0}$ yields the scattering density $\rho_1$ of the first fractal as the contrast match point, and the quantity $n_{1}V_{1}^{2}/L_{1}^{\tau_{1}}$. Plotting  $\sqrt{a_{2}}$ versus $\rho_{0}$ allows us to distinguish between the types \emph{I} and \emph{II}: if $a_2$ is independent of $\rho_0$, we deal with a structure of type $I$, otherwise we deal with a structure of type $II$. In the latter case, the parameters $\rho_{2}$ and $n_{2}V_{2}^{2}/L_{2}^{\tau_{2}}$ can be found as well. Thus, the Stuhrmann variation methods \cite{stuhrmann70} is applied not to the scattering intensity at zero angle but to the coefficients $a_1$ and $a_2$.

We emphasize that the suggested multi-phase model neither contradicts nor denies the results of two-phase models such as Beaucage's model \cite{beaucage95}, but our model completes and supplements them. Indeed, the contrast variation method suggests a clear and transparent criterion to understand whether a two-phase model is relevant to a specific power-law crossover, observed experimentally, or not. If the crossover point $q_c$ depends on the surrounding scattering-length density then it cannot be explained in the framework of a two phase model. Note that the crossover position can be very sensitive to the contrast variation (see the discussion at the end of Sec.~\ref{sec:application}).

As is discussed in Introduction, the presence of the power-law behaviour in the scattering intensity is sufficient but not necessary condition for the existence of fractals in the sample.  The power-law behaviour can be `occasional' and  appears due to polydispersity in a rather narrow range in momentum space. We emphasize that even in this case the above analysis allows us to distinguish between structures of types \emph{I} and \emph{II}.

In the previous papers \cite{chernyJACR10,chernyJSI10}, simple polydisperse models of mass fractal are considered, for which scattering intensities are similar to that of shown in Fig.~\ref{fig:fig1}a. Here we develop the model, which is quite close to real physical structures with rather complex distributions of scattering length density and whose behaviour is different essentially from the `standard' one shown in Fig.~\ref{fig:fig1}a. Then the contrast variation experiments are needed to reveal the real space structure of investigated systems.

\acknowledgements
The authors are grateful to Jos\'e Teixeira for valuable discussions. The authors acknowledge support from JINR–--IFIN-HH projects and from JINR
grant No. 13-302-02.

\bibliography{sasmfs}

\end{document}